\documentclass[prb,showpacs,twocolumn,preprintnumbers,amsmath,amssymb,floatfix]{revtex4}
\usepackage[dvips]{graphicx}
\usepackage[latin1]{inputenc}
\usepackage{stmaryrd}

\begin{document}
\draft
\title{NMR investigation of vortex dynamics in Ba(Fe$_{0.93}$Rh$_{0.07}$)$_2$As$_2$ superconductor}
\author{L. Bossoni,$^{1,2}$ P. Carretta,$^{1}$ A. Thaler,$^{3}$ P.C.
Canfield$^{3}$.}
\address{$^{1}$ Department of Physics ``A. Volta,'' University of
Pavia-CNISM, I-27100 Pavia, Italy}
\address{$^{2}$ Department of Physics ``E. Amaldi,'' University of Roma Tre-CNISM, I-00146 Roma, Italy}
\address{$^{3}$ Ames Laboratory US DOE and Department of Physics and
Astronomy, Iowa State University, Ames, IA 50011, USA}

\widetext

\begin{abstract}
$^{75}$As NMR spin-lattice relaxation ($1/T_1$) and spin-echo decay ($1/T_2$)
 rate measurements were performed in a single crystal of
Ba(Fe$_{0.93}$Rh$_{0.07}$)$_2$As$_2$ superconductor. Below the
superconducting transition temperature $T_c$, when the magnetic
field $\mathbf{H}$ is applied along the $c$ axes, a peak in both
relaxation rates is observed. Remarkably that peak is suppressed
for $\mathbf{H}\perp c$. Those maxima in $1/T_1$ and $1/T_2$ have
been ascribed to the flux lines lattice motions and the
corresponding correlation times and pinning energy barriers have
been derived on the basis of an heuristic model. Further information on the flux lines motion was derived from the
narrowing of $^{75}$As NMR linewidth below $T_c$ and found to be consistent with that obtained from $1/T_2$ measurements. All the experimental
results are described in the framework of thermally activated
vortices motions.
\end{abstract}

 \maketitle

\draft

\narrowtext

\section{INTRODUCTION}
The recent discovery of iron-based superconductors~\cite{Kam2008}
was welcomed by the scientific community, because it was supposed
to answer  the still open questions regarding the pairing
mechanism in high temperature superconductors. However the
multiplicity of controversial experimental results suggests that a
unique description of the superconducting properties is far from
being reached. Among the still debated fundamental topics, e. g.
the order parameter symmetry,\cite{Ding2008,Kondo2008,Szabo2009,Mu2009,Terashima2009} the
nanoscopic coexistence of magnetism with superconductivity,\cite{Sanna2009,Laplace2009} the role of antiferromagnetic spin
fluctuations in the pairing mechanism,\cite{Fang2011,Mazin2008,Mazin2008_2} one fixed point is
represented by the study of the flux lines lattice (FLL).\cite{Blatter1994} In fact,
the study of the magnetic field ($H$)-temperature ($T$) phase
diagram  of iron-based superconductors have immediately attracted
lot of interest owing to their extremely high upper critical
fields,\cite{Ni2008,Ni2009} which in some cases reach values even larger than those of high $T_c$ superconductors.

Most of the theories aiming at describing the FLL properties are
based on a regular arrangement of vortices.\cite{Abrikosov1957} However, in real
crystals this is far from being the case, because crystal defects,
such as dislocations or inclusions, usually act as pinning centers
preventing the vortices from having a regular arrangement or from
moving freely under the action of an electric current. Since these
dynamics lead to dissipative effects the study of the pinning
potential is of major importance for the technological
applications of superconductors. On the other hand, the
understanding of the different phases developing in the magnetic
field-temperature phase diagram of a superconductor and the
modelling of the different dynamical regimes give rise to fundamental
questions.~\cite{Gammel} A technique which offers the possibility
of studying the FLL dynamics from a microscopic point of view is
nuclear magnetic resonance (NMR). In the past years a fruitful study was
performed in the cuprates~\cite{Rigamonti1998,Carretta1992,Carretta1993} and showed that
the linewidth and the spin-lattice relaxation times were effective
markers  of the vortex dynamics. Moreover, these two quantities
provide complementary information since the linewidth narrowing is
sensitive to the magnetic field fluctuations along the direction
of the external field, while the spin-lattice relaxation time
is sensitive to the transverse fluctuations.
Furthermore, being the nuclei local probes they are sensitive to flux lines
excitations at all wave-vectors,~\cite{Torchetti} at variance with macroscopic
techniques, as the AC susceptibility for example, which are sensitive just
to the long wavelength excitations.\cite{Prando2011}

Thanks to the works performed in the cuprates we know now that in
very anisotropic superconductors vortices can be considered as independent two-dimensional isles called "pancakes" which undergo diffusive thermal motions.\cite{Anderson,Tinkham}
Bearing this in mind, and looking at the structural similarities
between cuprates and pnictides, some obvious questions arise: is
it still possible to detect the vortices thermal dynamics in
iron-pnictides with NMR? What is the vortices structure in the new
iron-based compounds? Are vortices 2D uncorrelated islands or
rather three-dimensional structures? In order to answer at least
part of these open questions we performed an NMR study of the
superconducting state of
Ba(Fe$_{1-x}$Rh$_x$)$_2$As$_2$ superconductor with x$\sim 0.07$. We measured both the spin-lattice ($1/T_1$) and spin-echo
($1/T_2$) relaxation rates of the $^{75}$As nuclei, together with
the Knight Shift and the NMR linewidth, at two different field
intensities (7 T and 3 T) and orientations ($\mathbf{H}\parallel$
or $\perp c$). The study of these quantities evidences the
presence of low-frequency dynamics that we interpreted in the
light of FLL motion and, accordingly, we derived a quantitative
description of the vortex motions, namely the temperature
dependence of the correlation time and of the pinning potential at
different magnetic fields. In the present work we will concentrate
solely on the superconducting properties, while the discussion of
the normal state will be presented in a future study.

\section{TECHNICAL ASPECTS AND EXPERIMENTAL RESULTS}

\begin{figure}[!h]
\includegraphics[height=8cm,width=8cm, keepaspectratio]{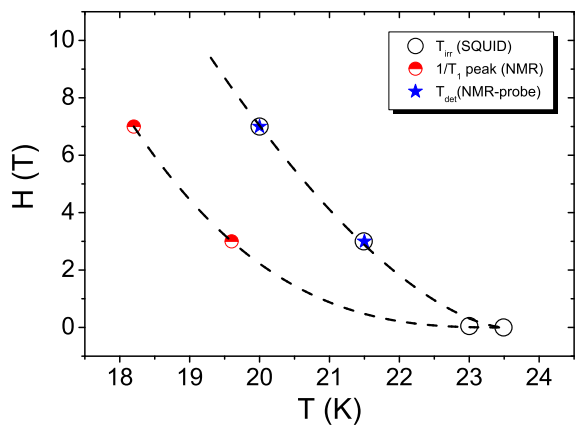}
\caption{The irreversibility temperature measured with a DC SQUID
magnetometer (open circle) is compared with that derived from the
detuning of the NMR probe (blue stars). The red circles refer to
the temperature of the peaks in $1/T_1$. The dotted lines are guide to the eye.}
\label{Tirr}
\end{figure}

$^{75}$As NMR measurements were performed on a flat 0.8 x 5 x 7
mm$^3$ parallelepiped shaped crystal of
Ba(Fe$_{0.93}$Rh$_{0.07}$)$_2$As$_2$ with the $c$ axis along the
shortest side. The sample was grown by  self-flux method according
to a procedure reported in Ref.\onlinecite{Ni2009}. The phase
diagram of Rh-doped compounds shows many similarities with that of
Co-doped compounds and the maximum estimated transition
temperature is about $23$ K~\cite{Ni2008} for the optimally doped
$x\simeq 0.07$ system. For such Rh concentration both the structural and
antiferromagnetic phase transitions are suppressed. To provide a
first characterization of the crystal we measured the field cooled
(FC) and zero field cooled (ZFC) magnetization by means of a
Quantum Design MPMS-XL7 Superconducting QUantum Interference
Device (SQUID) magnetometer. The irreversibility line was estimated looking at the temperature where the ZFC curve departs from the FC one, as in Ref.\onlinecite{Lascialfari}. This is the temperature where the magnetization is sensitive to a change in the dynamics of the FLL. On the other hand the detuning of the NMR probe~\cite{Carretta1992} is a higher frequency measurement, which is found to be consistent with what is observed in static field measurements (Fig. \ref{Tirr}).

\begin{figure}[!h]
\includegraphics[height=9cm,width=8cm, keepaspectratio]{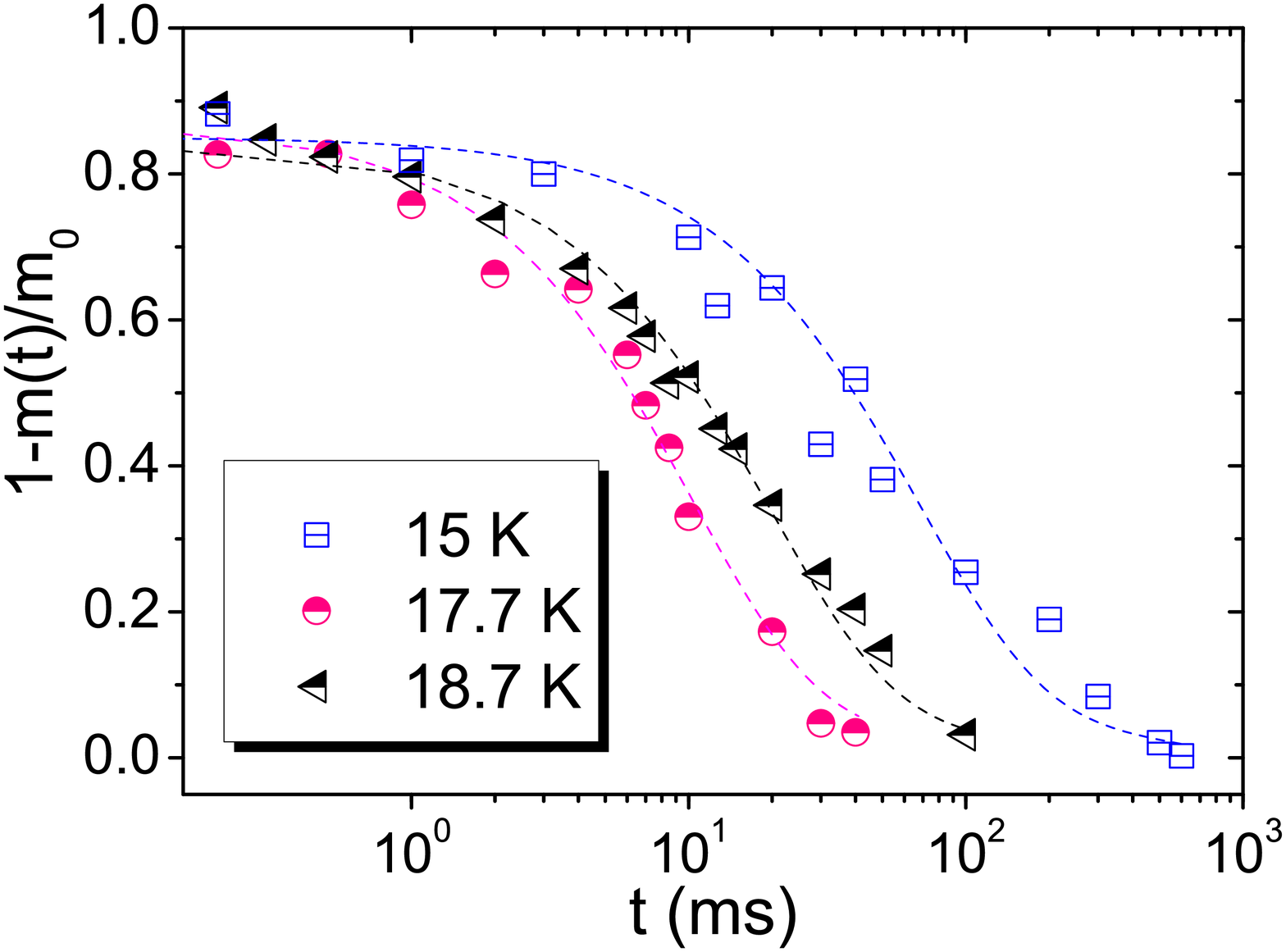}
\caption{The recovery curves for three different temperatures are
shown, for $\mathbf{H} \parallel c$, at 7 T. The blue squares
refers to the 15 K data, while the pink circles are taken at 17.7
K, in correspondence with the peak in $1/T_1$, and the black
triangles refer to 18.7 K. The dotted coloured lines are the best
fits according to Eq. 1.} \label{recov}
\end{figure}

\begin{figure}[!h]
\includegraphics[height=9.3cm, width=8cm, keepaspectratio]{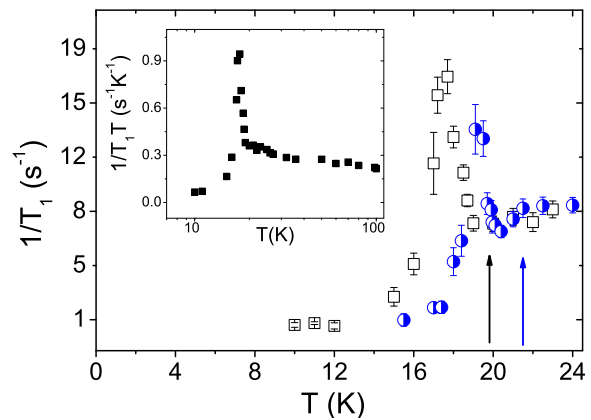}
\caption{The spin-lattice relaxation rate, measured at 7 T (open squares) and 3 T (blue circles), for $\mathbf{H}\parallel c$ is reported. The inset shows the $1/T_1 T$ data at 7 T both in the superconducting and normal phase. The arrows show the temperature of the detuning of the NMR probe at the two fields: the blue arrow stands for 3 T and the black arrow for 7 T.}
\label{2fields}
\end{figure}

\begin{figure}[!h]
\includegraphics[height=9cm,width=8cm, keepaspectratio]{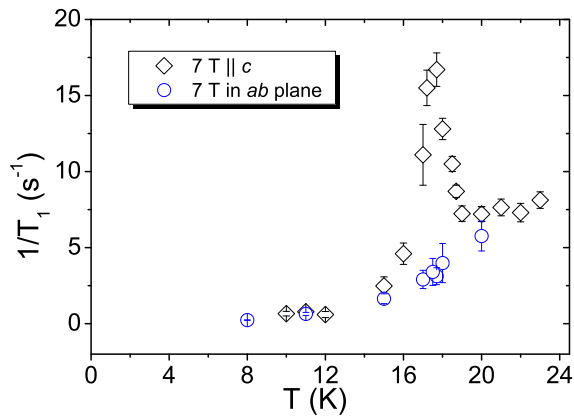}
\caption{The spin-lattice relaxation rate, measured at 7 T, in $\mathbf{H}\parallel c$ geometry (black diamonds) and $\mathbf{H}\perp c$ geometry (blue circles) is shown. A neat difference for the two field orientations is found in the 16-19 K range. Data, in $\mathbf{H} \perp c$ geometry, have been normalized by a value 1.55 to match the value of $1/T_1$  for $\mathbf{H}\parallel c$, at $T_c$ thus revealing an anisotropy of the hyperfine tensor.}
\label{cfr_orient}
\end{figure}

The NMR measurements were performed by using standard
radiofrequency pulse sequences. The spin-lattice relaxation time
$T_1$ was measured by means of a saturating recovery pulse
sequence at two different magnetic fields $H =$ 7 T and 3 T. The
recovery of the nuclear magnetization $m(t)$ was found to follow
the relation \cite{Fukazawa2008,Simmons1962}:
\begin{equation}
1-m(t)/m_0=0.1 e^{-t/T_1}+0.9 e^{-6t/T_1}
\end{equation}
expected for a nuclear spin $I=3/2$ in the case of a magnetic
relaxation mechanism (see Fig. \ref{recov}). In the normal phase
$1/T_1T$ shows a temperature independent behavior, as expected for
a weakly correlated metal (see the inset into Fig.
\ref{2fields}).~\cite{Slicther} By decreasing the temperature
below $T_c$ we observed a well-defined peak in $1/T_1$ for
$\mathbf{H}\parallel c$. The peak temperature decreased by
increasing the magnetic field intensity (see Fig. \ref{2fields}).
Remarkably when $\mathbf{H}\perp c$ the peak in $1/T_1$ disappears
(see Fig. \ref{cfr_orient}). At lower temperatures $1/T_1$
decreases exponentially and it is only weakly dependent on the
magnetic field orientation.


\begin{figure}[!h]
\includegraphics[height=9cm,width=8cm, keepaspectratio]{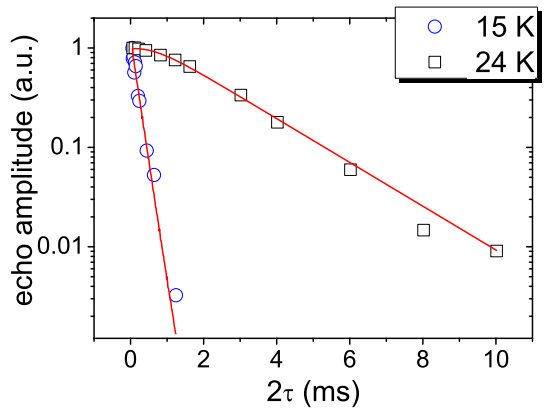}
\caption{The figure shows two echo decays as a function of
2$\tau$, below $T_{det}$: the black squares refer to 24 K and the
blue circles refer to 15 K, below the superconducting transition
where the FLL is still dynamic. From the figure one can notice
that the echo decay functional form changes while the temperature
decreases. The red curves are the best fits according to Eqs. 10
and 11.} \label{echi}
\end{figure}

The transverse relaxation time $T_2$ was measured by recording the
decay of the echo after a $\pi/2-\tau -\pi$ pulse sequence as a
function of the delay $\tau$. Since the functional form of the
decay changes with temperature (Fig. \ref{echi}), as will be
discussed subsequently, in order to compare the data over the full
temperature range we defined $T_2$ as the time where the echo
amplitude decreases by 1/e.

\begin{figure}[h!]
\includegraphics[height=9cm,width=8cm, keepaspectratio]{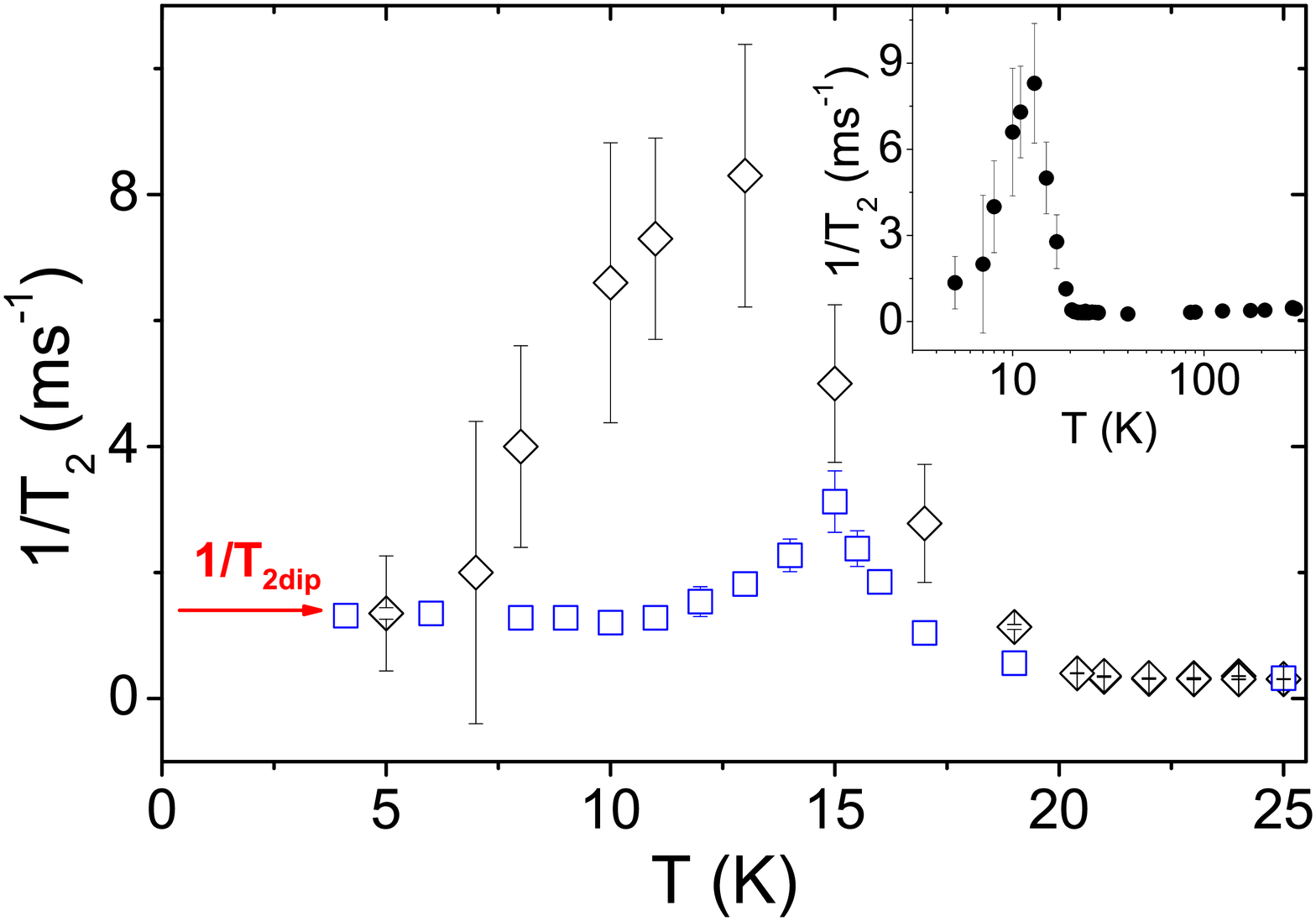}
\caption{The figure shows the spin echo decay rate measured at 7 T. A peak around 12 K is found for $\mathbf{H}\parallel c$ (black diamonds) while it strongly decreases, for $\mathbf{H}\perp c$ (blue squares), and shifts towards higher temperatures. The red arrow indicates the ab-initio value for $1/T_2$ given by the dipolar sums. The inset shows the spin echo decay rate for $\mathbf{H}\parallel c$ up to room temperature.}
\label{T2}
\end{figure}

In the normal phase $1/T_2$ shows an activated temperature
dependence whose origin will be discussed elsewhere. Below $T_c$
we observed a marked increase in 1/T$_2$ giving rise to a peak
around 12-13 K, for $\mathbf{H}\parallel c$. We note that Oh et \textit{al.} found a similar behaviour in their data referred to the 7.4\% Co-doped single crystal.~\cite{Oh2011} Nevertheless we note that the compound is different though $1/T_2$ data are quite similar: they observed a peak in $1/T_2$ around 15 K while we observed it around 12-13 K. For $T\rightarrow 0$
the spin echo decay rate is found to reach the value derived from
nuclear dipolar lattice sums. \\
Similarly to what was observed for
$1/T_1$, also $1/T_2$ peak gets significantly reduced for
$\mathbf{H}\perp c$ (Fig. \ref{T2}).

\begin{figure}[h!]
\begin{minipage}[t]{.53\textwidth}
\includegraphics[width=0.8\textwidth]{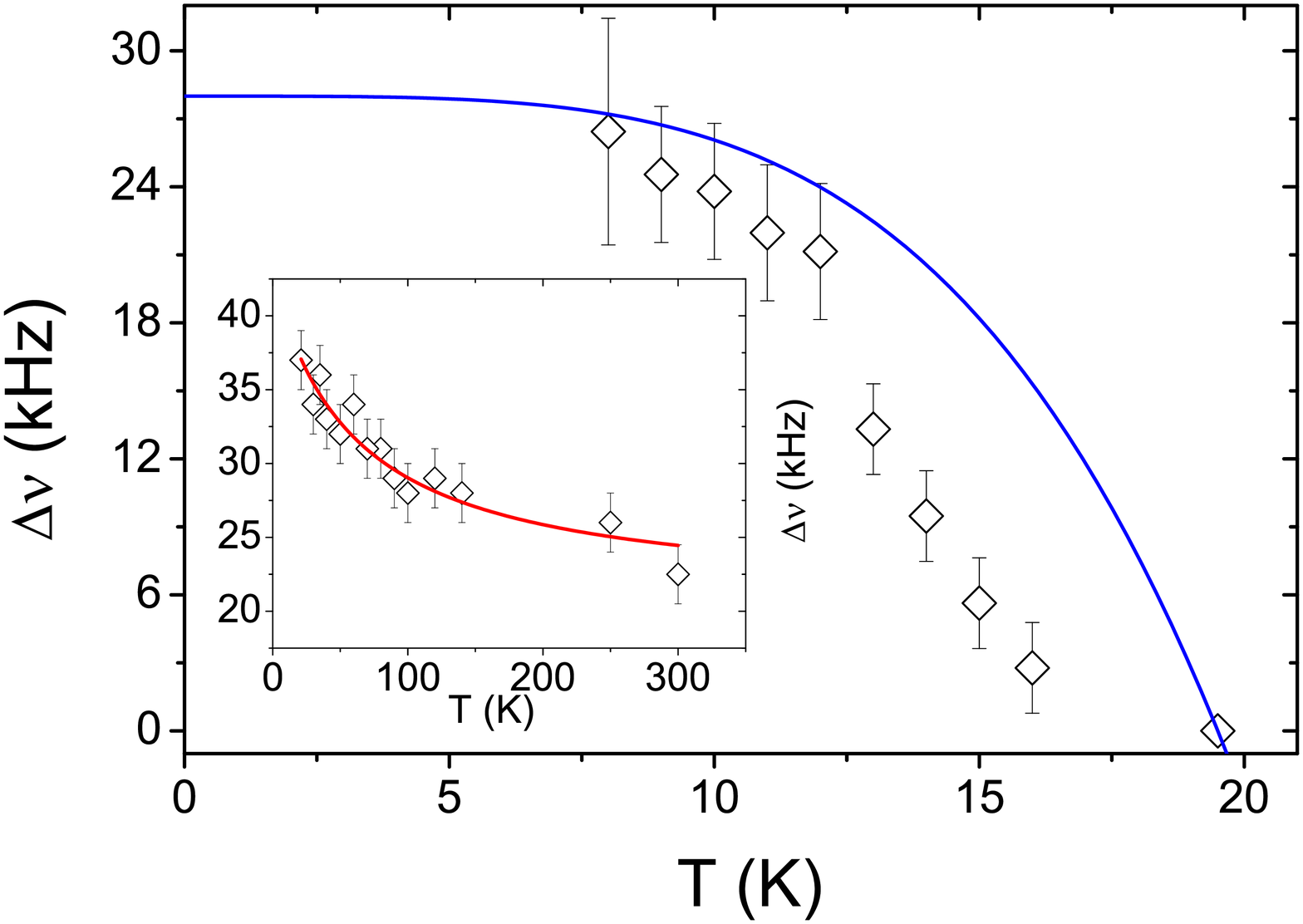}
\end{minipage}
\vspace{9mm}
\begin{minipage}[t]{.53\textwidth}
\includegraphics[width=.8\textwidth]{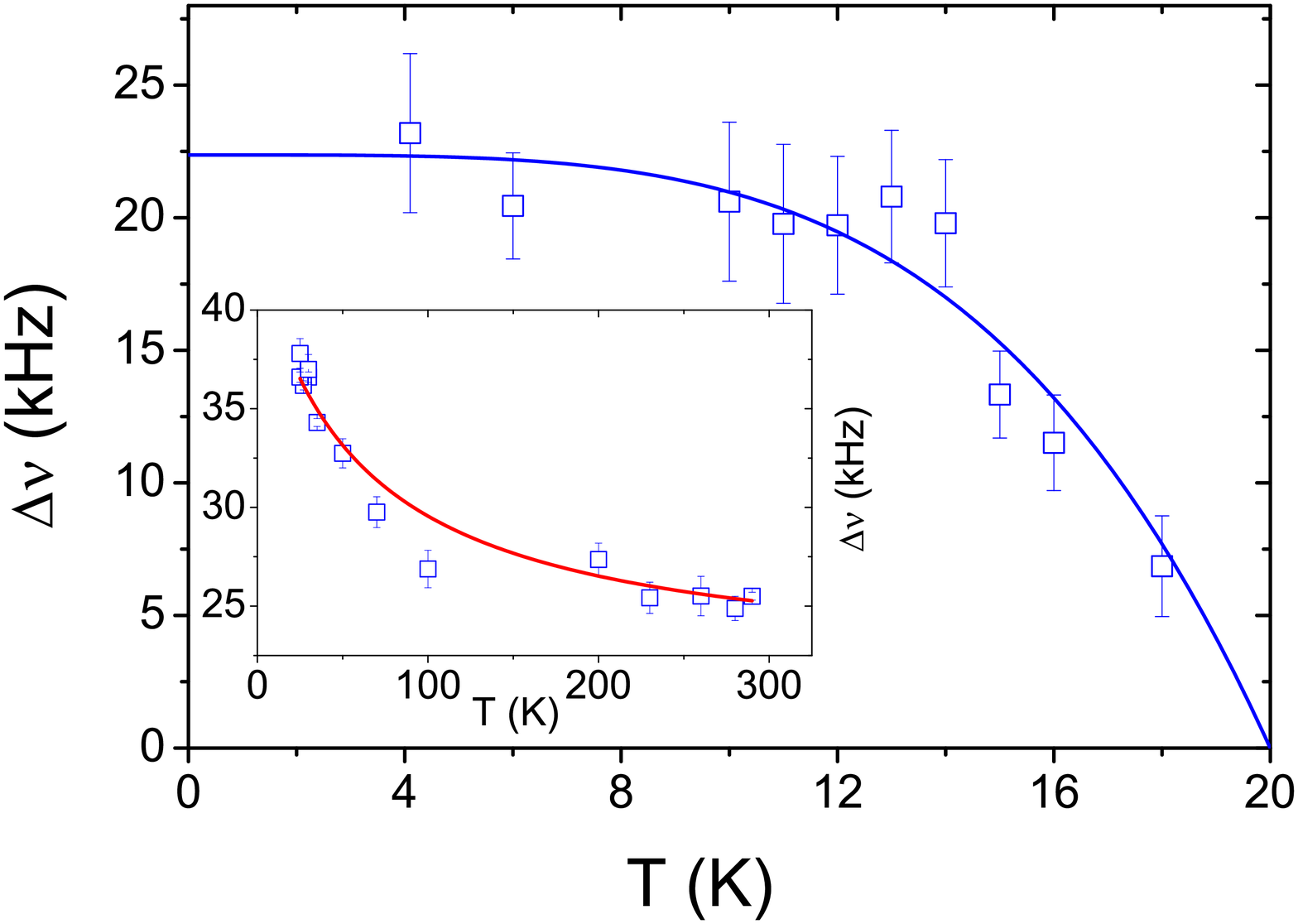}
\end{minipage}
\caption{The figure at the top shows the FWHM at 7 T, for $\mathbf{H}\parallel c$, after the Curie-Weiss correction (see Eq. \ref{CW}). An effect of narrowing occurs just below $T_c$: the experimental data deviate from the London two-fluid model (blue line).
The figure at the bottom shows the FWHM at 7 T, for $\mathbf{H}\perp c$, after the Curie-Weiss correction. Here the extra-broadening can be fitted by the London two-fluid model, up to $T_c$. In the insets the Curie-Weiss behavior observed for the raw data is shown.}\label{LW}
\end{figure}

The NMR spectrum was determined from the Fourier transform of half
of the $^{75}$As echo signal, while below $T \simeq 13$ K, when the
line became too broad, the spectrum was derived by sweeping the
irradiation frequency. The full width at half maximum (FWHM) was
determined by a Gaussian fit. In the normal state the
linewidth increased on cooling following a Curie-Weiss
trend (Fig. \ref{LW}), probably due to the presence of
paramagnetic impurities. The impurities cause the appearance of a staggered magnetization and a broadening of the NMR line. On the other hand, the average magnetic field is only weakly affected, so we do not expect an extra-contribution to the shift.~\cite{Alloul}
After subtracting this impurity-dependent contribution $\Delta \nu_{NP}$ from the raw data by using the
relation
\begin{equation}
\Delta \nu(T) \simeq \sqrt{\Delta \nu (T)_{raw}^2-\Delta \nu_{NP}^2} \label{CW}
\end{equation}
we observed that below $T_c$ an extra-broadening induced by the
presence of the flux lines lattice appears (Fig. \ref{LW}).
The impurity-dependent contribution was found to be well described by the Curie-Weiss relation, for both the sample orientations:
\begin{equation}
\Delta \nu_{NP}(T)=\frac{C}{T-\theta}+A
\end{equation}
By assuming the value of $\theta=-60$ K, as found by the fitting procedure we obtained the following results: for the $\mathbf{H}\parallel c$ case the fit gave $C=1319 \pm 118$ kHz K and $A=20.8 \pm 1$ kHz, while for the perpendicular geometry the fit gave $C=1264 \pm 50$ kHz K and $A=23.4 \pm 0.1$ kHz.


\begin{figure}[!h]
\includegraphics[height=10cm, width=8cm, keepaspectratio]{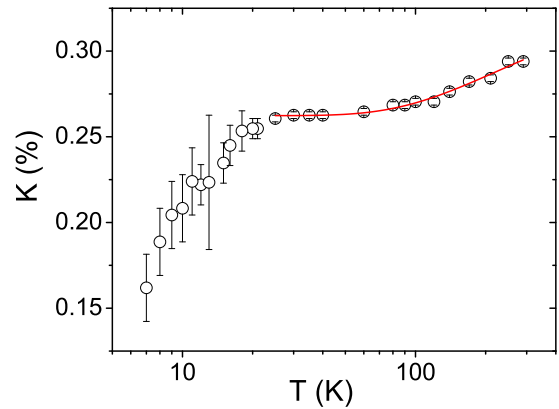}
\caption{The figure shows the Knight Shift at 7 T, for $\mathbf{H}\parallel c$ with an Arrhenius-like fitting curve (solid line) in the normal phase.}\label{Ks}
\end{figure}

It has to be noticed that the superconducting state affects not
only the $^{75}$As NMR linewidth but also the NMR shift. Above
$T_c$, in the normal phase, the NMR shift shows an activated
behavior, as observed also for the Co-doped
BaFe$_2$As$_2$.~\cite{Oh2011} The experimental data (Fig.
\ref{Ks}) can be fit with an activated Arrhenius law:
$y=A+B\exp(-D/T)$, yielding $A=0.26$ \%, $B= 0.071$ \% and $D=225$
$\pm 22$ K,  for $\mathbf{H}\parallel c$, in good agreement with
the values found in Ref.\onlinecite{Oh2011}. Below $T_c$ the
shift starts to decrease as expected for a singlet state
pairing.\cite{Yosida} In the superconducting phase the NMR shift $K(T)$ can be
assumed to result from three contributions:
\begin{equation}
K(T)=K_{spin}(T)+K_{FL}(T)+K_{TI}
\end{equation}
where $K_{spin}(T)$ is the spin-dependent part, which vanishes for
$T\rightarrow 0$, $K_{FL}(T)$ is the diamagnetic correction due to
the vortex lattice, and the last term contains all the temperature
independent contributions (chemical shift, orbital terms, etc...).\cite{Mac1976}
Owing to the line broadening and to the reduction in the
radio-frequency penetration depth, the accuracy in the estimate of
$K(T)$ decreases below $T_c$ and does not allow us to draw
convincing conclusions on the symmetry of the order parameter.

Taking into account the quadrupolar shift of the central line, in the transverse geometry, we estimated a quadrupolar frequency $\nu_Q $(100 K) $\sim 1.5$ MHz, very close to the one found in the parent compound.~\cite{Fukazawa2008}

\section{DISCUSSION}

As previously mentioned, here we will not discuss the normal state
properties but rather we shall concentrate on the superconducting
phase. Let us first consider the behaviour of the spin-lattice
relaxation rate which is characterized by a well defined peak for
$\mathbf{H}\parallel c$ below $T_c$. In passing, we note that in Co-optimally doped compound,~\cite{Ning2008} no peak was observed in $1/T_1$, below $T_c$. On the other hand Laplace et \textit{al.},~\cite{Laplace2009} in a 6\% Co-doped BaFe$_2$As$_2$,
found a peak in $1/T_1$ just below $T_c$ and an enhancement in $1/T_1T$ at higher temperature due to spin density wave correlations.
The peak we found below $T_c$ is not expected to be a
Hebel-Slichter peak~\cite{HS1959} since the majority of the
experimental and theoretical results point towards an extended
s$^{\pm}$-wave
pairing,\cite{Ding2008,Szabo2009,Mu2009,Terashima2009,Parker2008} where that
feature is expected to be absent. Furthermore if it was a coherence
peak the data would be described, below $T_c$, by $1/T_1\sim e^{-\Delta/T}$ with $\Delta$ the
superconducting gap. By fitting the data one obtains $\Delta\simeq
200$ K $>> 3.5 k_BT_c$, the value expected for the superconducting
order parameter.~\cite{Nakayama2009} Finally, the striking suppression of the
peak for $\mathbf{H}\perp c$ can hardly be reconciled with the
small anisotropy of the electron spin susceptibility found in
those materials. Hence that maximum in $1/T_1$ just below $T_c$
should not be associated with the electron spin dynamics but,
given the similarities with the behavior found in
HgBa$_2$CuO$_{4+\delta}$~\cite{Suh1996} and
YBa$_2$Cu$_4$O$_8$~\cite{Corti1996} cuprates, it is tempting to
associate $1/T_1$ peak to the FLL dynamics.

In order to analyze the experimental results one can start from
the basic modelling of FLL in strongly anisotropic superconductors:~\cite{Brandt1991} the vortices enter the sample in form of
quasi-two dimensional pancakes, lying in the FeAs planes. Owing to
the thermal excitations they move out of their equilibrium
positions by means of random motions, which can be hindered by the
pinning centers. Differently from cuprates, which exhibit a very
high anisotropic ratio $\gamma= \xi_{ab}/\xi_{c}$, with $\xi_{ab}$
and $\xi_c$ the in-plane and out of plane coherence lengths, the
Ba122 superconductors show $\gamma \sim $2-4 varying with
temperature.~\cite{Ni2008} This suggests to describe the flux lines
not as completely uncorrelated pancakes, but rather as a stack of
correlated islands. Still, since the estimated correlation length
$\xi_c$ is of the order of the inter-layer distance $s$,~\cite{Rotter2008} namely $2\xi_c \simeq s \simeq 6$\AA\ , FeAs planes can
be considered as weakly coupled superconducting layers.
Accordingly, when $\mathbf{H} \perp c$ the flux lines are
preferentially trapped between the planes and the FeAs plane
boundaries act as pinning centers, a
well known effect in layered superconductors.
These intrinsic pinning centers hinder the dynamics and yield the observed
suppression in the $1/T_1$ peak for $\mathbf{H} \perp c$.

In order to understand the shift of the $1/T_1$ peak
upon increasing $H$ we first recall that $1/T_1$ probes the
spectral density $J(\omega_L)$ at the nuclear Larmor frequency
$\omega_L$, namely
\begin{equation}
\frac{1}{T_{1}}=\frac{\gamma^{2}}{2}\int<h_{\rho}(t)h_{\rho}(0)>e^{-i\omega_L
t}dt \label{1}
\end{equation}

with $h_{\rho}$ the magnetic field component perpendicular to
$\mathbf{H}$ and $\gamma= 2\pi\times  7.292\times 10^6$ rad/T the
gyromagnetic ratio of the $^{75}$As nucleus. Then the field
dependence of the peak in $1/T_1$ can be qualitatively understood
by considering that at the peak temperature the characteristic
frequency for FLL motions is close to $\omega_L$. When the
magnetic field increases, $T_{c}$ decreases and so does $T_{irr}$,
hence the FLL dynamics remain fast over a broader temperature
range and the maximum in $1/T_1$ is observed at lower temperature
(Fig. 1). It is noticed that the peak in the spin-lattice
relaxation rate appears just below the irreversibility
temperature, in contrast with what was found in the cuprates,
where it is well below the irreversibility line, suggesting a
higher FLL mobility in these latter compounds.~\cite{Corti1996}

To give a quantitative description of the peak we started from the
equation \ref{1}. Let us first assume that the vortex
fluctuations are basically two-dimensional (2D), that take place in
a spatial range smaller than the inter-vortex
distance~\cite{Pavuna} $l_{e}=\sqrt{{2}/{\sqrt{3}}}\sqrt{{\Phi
_{0}}/{H}}$ (for a triangular FLL) and that they move by Brownian
motions~\cite{Corti1996,Suh1996} described by a diffusive-like
correlation function $g_{1}(t)=\exp(-D_{\perp}q_{\perp}^{2}t)$,
$D_{\perp}$  being the diffusion constant of the motion taking
place in the $ab$ plane. Then
$\tau_c(q_{\perp})=1/D_{\perp}q_{\perp}^{2}$ plays the role of a
q-dependent correlation time for the collective vortex motions. By
summing over all collective in-plane excitations up to a cut off
wave-vector $q_{m}= (1/l_e)(8\pi^3/3)^{1/4}$ Suh et
\textit{al.}~\cite{Suh1996} found the spectral density
\begin{equation}
J(\omega_{L})=\tau_m\ln
\left[\frac{\tau_m^{-2}+\omega_{L}^{2}}{\omega _{L}^{2}}\right]
\label{JJ}
\end{equation}
where the average correlation time is $\tau_{m}=1/D_{\perp}q_{m}^{2}$. For the temperature
dependence of $\tau_m$ it is reasonable to assume an activated
form $ \tau_m (T)=\tau_0 \exp(U/T) $, where $U$ is an average
pinning energy barrier and $\tau_0$ stands for the correlation
time in the infinite temperature limit. Accordingly an activated
temperature dependence of the spectral density at the Larmor
frequency and then of $1/T_1$ are observed for $T\rightarrow 0$.
The best fits of the data according to this 2D vortex model are
reported in Fig. 8. It is noticed that the fit is not
fully satisfactory.

\begin{figure}[h!]
\begin{minipage}[t]{.53\textwidth}
\includegraphics[width=0.8\textwidth]{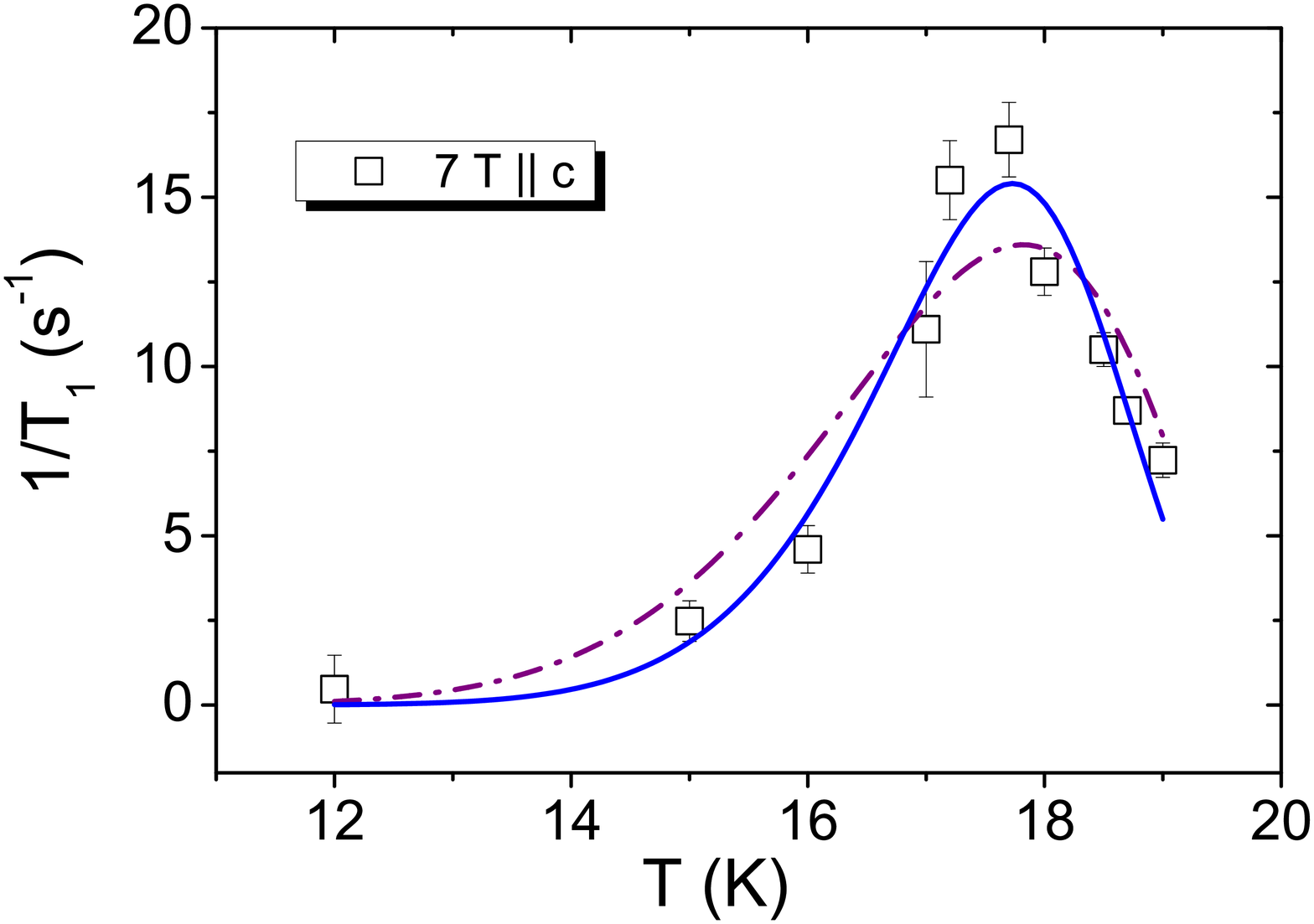}
\end{minipage}
\vspace{9mm}
\begin{minipage}[t]{.53\textwidth}
\includegraphics[width=.8\textwidth]{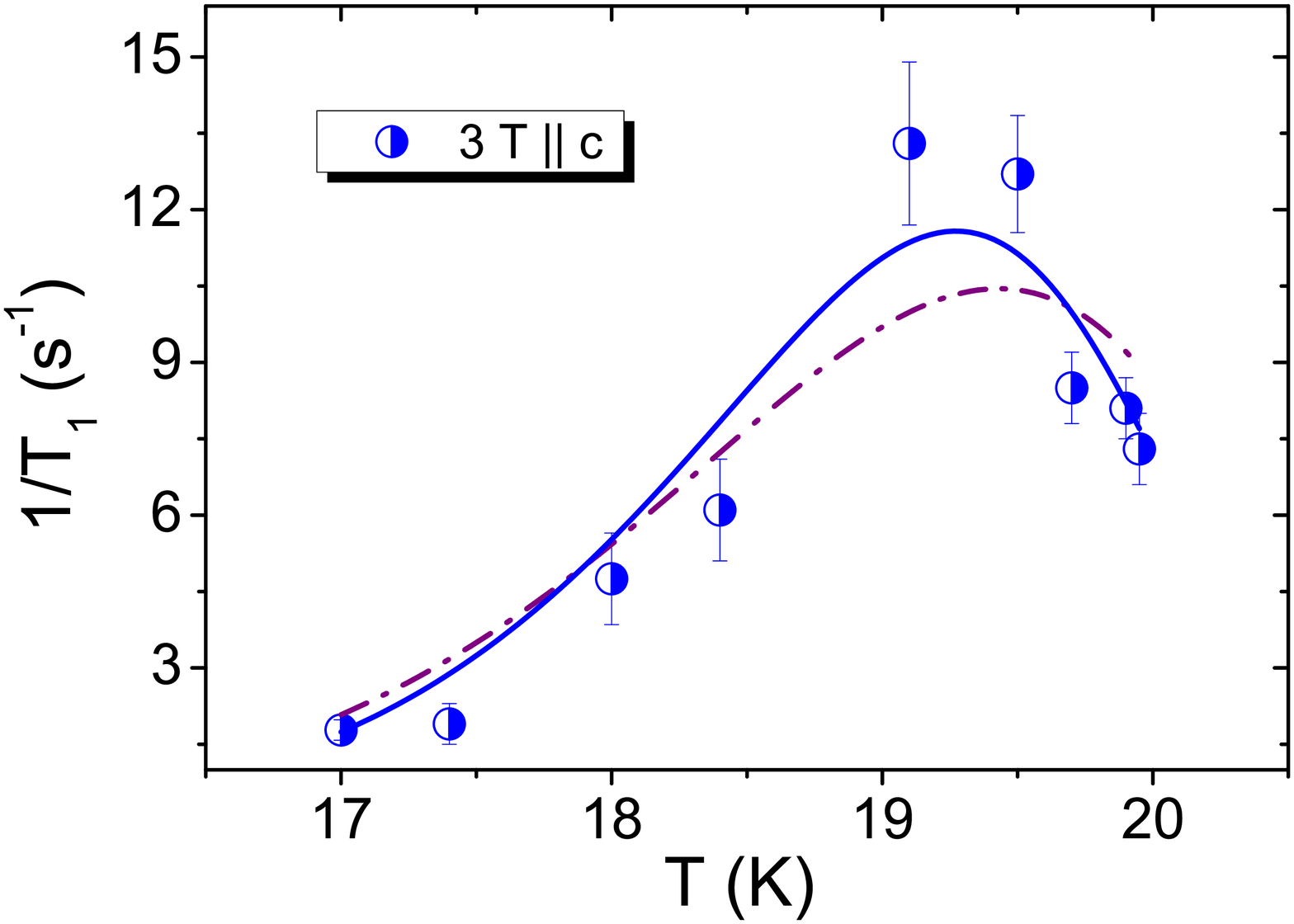}
\end{minipage}
\caption{The figure at the top shows the spin-lattice relaxation rate at 7 T for $\mathbf{B} \parallel c$, while the one at the bottom shows the spin-lattice  relaxation rate at 3 T in the same geometry. The fitting curves are given by the 2d - uncorrelated pancakes model, deriving from the correlation function $g_1(t)$ (dash-dotted line) and the correlated vortices model, deriving from the correlation function $g_2(t)$ (solid line). In both cases the second model shows the best agreement with the experimental data. }
\label{fit_T1}
\end{figure}

On the other hand, as previously pointed out, the low anisotropy
of BaFe$_2$As$_2$ compounds suggests that significant vortex
correlations along the $c$ axes are present in
Ba(Fe$_{0.93}$Rh$_{0.07}$)$_2$As$_2$. Thus, the flux lines have to
be considered as stationary waves oscillating in between the
pinning centers. In order to take into account this effect we
introduced empirically a modulation in the amplitude of the
correlation function characterized by a wavelength $\lambda$ which
has an upper bound given by $\lambda_c$, the London penetration
depth along the $c$ axes. Then one can write
$g_2(t)=\exp(-D_{\perp}q_{\perp}^{2}t) \cos
\left({z}/{\lambda}\right)$. Now if we recall the form of the
longitudinal field correlation function\cite{Corti1996}
\begin{eqnarray}
<h_{\rho}(0)h_{\rho}(t)>&=& \frac{\Phi_0^2 s^2 }{4 \pi \lambda_c^4}<u^2> \frac{1}{\xi^2}\frac{1}{l_e^2\sqrt{3}}g_2(t)
\label{h}
\end{eqnarray}
and taking the root mean square amplitude of the vortex core fluctuation with respect to equilibrium position \cite{Brandt1992,Clem1991}
\begin{eqnarray}
<u^2> \simeq \frac{\sqrt{2\pi \sqrt{3}}}{\Phi_0^2}\lambda_c\lambda_{ab}l_e k_B T \nonumber
\end{eqnarray}
Eq. \ref{h} can be written:
\begin{eqnarray}
<h_{\rho}(0)h_{\rho}(t)>&=& \sqrt{\frac{3}{8\pi}}\frac{s^2 k_B T}{l_e}\frac{\lambda_{ab}(T)}{\lambda_{c}^3(T)\xi^2(T)}g_2(t) \label{2}
\end{eqnarray}
where the temperature dependence is evident. Taking the values for
the London penetration depth reported in the
literature,~\cite{Tanatar2009} the coherence lengths derived from
the $H_{c2}$ measurements, and their temperature dependence
according to the two-fluid model, we were able to reproduce fairly
well the temperature dependence of $1/T_1$, which indicates that
indeed a "3D-correlated-vortices" model is more appropriate to
describe Ba(Fe$_{1-x}$Rh$_x$)$_2$As$_2$ superconductors. The best
fits of the experimental results (Fig. \ref{fit_T1}) gives a value
of  $U$ $=322 \pm 66$ K for $H= 7$ T and $ U =470\pm 5$ K for $H=
3$ T. These values are similar in magnitude to those found in
YBCO-124,~\cite{Corti1996} nonetheless the quality of the fitting
procedure suggests that the vortices develop a three-dimensional
correlation. Before concluding this part we estimate the
root-mean-square amplitude of the transverse field fluctuations
$<h_e^2>$ which represents the ripple of the magnetic field
profile modulated by the flux lines dynamics. Infact given the
Eqs. \ref{1} and \ref{2}, ${1}/{T_1}$ can be written in this new
form ${1}/{T_1}= ({\gamma^{2}}/{2})<h_e^2>J(\omega_{L})$ from
which we obtained $h_e \sim 30-40$ Gauss at 7 T and $\sim$ 20
Gauss, at 3 T. We point out that these values are close to the low
temperature NMR full width at half maximum, as it has to be
expected.

While the spin-lattice relaxation rate has been considered one
of the most valuable microscopic probes of the FLL motion, not so
much effort has been devoted to the analysis of the spin echo decay time, mainly because its interpretation is not always
straightforward.~\cite{Suh1993,Recchia} As it has been
already pointed out in the superconducting state $1/T_2$ shows a
neat anisotropy: the peak found for $\mathbf{H} \parallel c$ is significantly
reduced and shifted in the $\mathbf{H} \perp c$ configuration. Moreover,
we point out that at low-temperature $1/T_2$ reaches the value
expected from the Van Vleck lattice sums.\cite{Oh2011,Sutirtha}
In fact, at low temperatures all
dynamics are frozen and the only process giving rise to the echo decay is
the nuclear dipole-dipole interaction. The peak observed around 10
K cannot be due to a time-dependent modulation of that interaction
since we do not expect such an anisotropic behavior, in that case. On the other
hand, given the similarity  with $1/T_1$ peak anisotropy, it is
likely that also $1/T_2$ peak arises from a low frequency vortex
dynamics. Nevertheless it should be emphasized that while $1/T_1$
measurements are sensitive to the fluctuations of the transverse
components of the magnetic field, $1/T_2$ is sensitive to the
longitudinal ones. In particular, it should be noticed that when
the vortices are strongly correlated along the $c$ axes, the flux
lines move rigidly and do not affect significantly the transverse
field components, while they do change the longitudinal ones. \cite{Clem1991}
Hence the information derived from those two types of
measurements can be complementary.

\begin{figure}[!h]
\includegraphics[height=8cm,width=9cm, keepaspectratio]{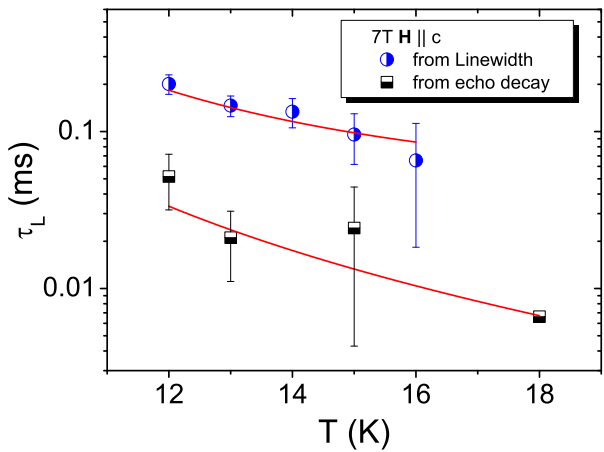}
\caption{The temperature dependence of the correlation time $\tau_L$ gives by echo decay time (black squares) is compared with the one derived by the linewidth analysis (blue circles) in the assumption of fast motions (see Eq. \ref{riga}). The red curves are the fitting of
the correlation time, according to an activated law.}
\label{tau}
\end{figure}

In order to analyze the temperature dependence of $1/T_2$ we need
an analytical expression for the spin echo decay. In principle we should start from a relation similar to Eq. \ref{h}, nevertheless here, for the sake of simplicity, we assumed an exponential correlation function for the longitudinal
field fluctuations
\begin{equation}
<h_l(0)h_l(t)>=<h_l^2>e^{-t/\tau_L}
\end{equation}
characterized by an average correlation time $\tau_L$. Correspondingly we have written the decay of the echo amplitude as a function of the delay $\tau$
between the $\pi/2$ and $\pi$ pulses in the echo sequence as \cite{Takigawa1986}
\begin{eqnarray}
M(2\tau)&=& M_0 e^{-2\tau/T_{2dip}^2} \times  M_2(2\tau)\\
M_2(2\tau)&=& e^{-\gamma^2<h_l^2>\tau_L^2 [2\tau/\tau_L+4\exp(-2\tau/\tau_L)- \exp(-2\tau/\tau_L)-3]}
\end{eqnarray}
where the first Gaussian term accounts for the nuclear
dipole-dipole contribution, while the second term describes the
low-frequency vortex motions. By fitting the data below $T_c$ we
were able to derive the temperature dependence of the longitudinal
correlation time (Fig. \ref{tau}). By decreasing the temperature
the FLL motion is supposed to go through different motional
regimes: from the fast motions ($\tau_L\ll T_2$) up to the very
slow motions ($\tau_L\gg T_2$), where the correlation time is so
long that we can consider the FLL to be frozen in the solid
state. If we fit the data above 11 K, where the peak in $1/T_1$ is
observed, we notice that $\tau_L$ follows an activated behavior
characterized by an activation barrier $U_L \simeq 50$ K much
lower than the one derived from $1/T_1$ (see. Fig \ref{tau}). Here
we refer just to the fast motion limit, since the fitting procedure is
not so much accurate at the low temperatures, because of the
reduced signal to noise ratio.

The correlation time derived from the spin-echo decay rate can be
suitably compared to that derived from the motional narrowing of
the NMR line. The latter can be derived, for the
$\mathbf{H}\parallel c$ case, following the standard approach
reported in Ref.~\onlinecite{Abragam}. In the fast motions regime,
namely $2\pi \overline{(\Delta \nu_R)^2}^{1/2}\tau_L<< 1$, with
$\overline{(\Delta \nu_R)^2}^{1/2}$ the square root of the rigid
lattice second moment, one has
\begin{equation} \Delta \nu \simeq
\tau_L \frac{\overline{(\Delta \nu_R)^2}}{2 \pi} \label{riga}
\end{equation}
By fitting the correlation time $\tau_L$ with an Arrhenius's law
we extracted a pinning energy barrier $U_L=48 \pm 3 $ K, which is
 consistent with the one derived from the spin echo decay
measurements. This is not surprising since both $T_2$ and $\Delta
\nu$ probe the longitudinal component of the local field fluctuation.

In order to understand why the energy barrier probed by $1/T_2$
and from the motional narrowing are smaller than the one derived
from $1/T_1$ measurements it should be pointed out that the
oscillations at wavevector $q_{\parallel} \rightarrow$ 0 do
contribute to the longitudinal field fluctuations but only weakly
to the transverse field excitations which are relevant in $1/T_1$,
at variance with the $q_{\parallel} \rightarrow 1/s$ modes, which contribute significantly to $1/T_1$. Hence,
our findings indicate that the energy cost to activate a certain
collective mode increases with increasing $q_{\parallel}$, where
the $\parallel$ subscript refers to the wavevector component
parallel to the magnetic field.

Upon cooling the crystal to the lowest temperatures we observed a
change in the line shape from Lorentzian to Gaussian, as it has to
be expected  when the correlation time gets longer than a few ms.
The Gaussian lineshape, instead of the asymmetric one expected for
a perfect triangular lattice, indicates the presence of lattice
distortions induced by randomly distributed pinning centers. In
this scenario one can only make an estimate of the London
penetration depth which can be compared with the one derived by transverse
$\mu$SR,~\cite{Williams2010} transport~\cite{Tanatar2009} and the
tunnel diode resonator measurements~\cite{Gordon2009} on a
similar 7.4\% Co-doped BaFe$_2$As$_2$ single crystal. These authors
report $\lambda_{ab}$ values between 200 and 217 nm. Following
Ref. \onlinecite{Brandt1988} we extracted:
\begin{equation}
\lambda_{ab}=\sqrt{\frac{2.36 \Phi_0 \gamma \sqrt{k}}{\Delta \nu}}
\end{equation}
where $\sqrt{k}= 0.04324$ depends on the lattice geometry and on the
magnitude of the applied field. Taking $\Delta \nu(T\rightarrow
0)\simeq 30$ kHz, for $\mathbf{H} \parallel c$ we extracted $\lambda_{ab} (0) \sim 226 \pm 9 $ nm, in agreement with the former results.


\section{CONCLUSIONS}

This work aimed at studying the thermally activated vortices motion,
by means of $^{75}$As microscopic probe. We found that in
the NMR relaxation rates and in the NMR linewidth there is
evidence of such motions, which is well supported by the
remarkably anisotropic behaviour of those quantities. We were
able to follow the dynamics of the vortices by measuring NMR
quantities which are sensitive to motions at different time
scales: the peak in the spin-lattice relaxation time is found when
the correlation time is about $10^{-7}-10^{-8}$ s, while the
$1/T_2$ maximum occurs at a slightly lower temperature, when the
correlation time is comparable to $1/T_2$, i.e. few ms. In the
temperature window between those peaks the motions are effective
and yield the motional narrowing of the NMR line. To observe a
line narrowing the correlation times must be smaller than the
inverse of the rigid lattice linewidth $ \sim 10^{-4}$ s.
Furthermore we pointed out that the temperature dependence of
$1/T_1$ around the peak suggests that the flux lines are formed by
strongly coupled vortices rather than nearly independent pancakes
diffusing in 2D, as it was found in the cuprates.

\section*{Acknowledgements}

We gratefully acknowledge Prof. A. Rigamonti for useful discussions, and M. Moscardini for his help in setting up the experiments.




\begin{references}

\bibitem{Kam2008} Y. Kamihara, T. Watanabe, M. Hirano, and H. Hosono, J. Am. Chem. Soc. \textbf{130}, 3296-3297 (2008).
\bibitem{Ding2008} H. Ding, P. Richard, K. Nakayama, T. Arakane, Y. Sekiba, A. Takayama, S. Souma, T. Sato, T. Takahashi, Z. Wang, X. Dai, Z. Fang, G. F. Chen, J. L. Luo, and N. L. Wang, Europhys. Lett. \textbf{83}, 47001 (2008).
\bibitem{Kondo2008} T. Kondo, A. F. Santander$-$Syro, O. Copie, C. Liu, M. E. Tillman, E. D. Mun, J. Schmalian, S. L. Bud'ko, M. A. Tanatar, P. C. Canfield, and A. Kaminski, Phys. Rev. Lett. \textbf{101}, 147003 (2008).
\bibitem{Szabo2009} P. Szabo, Z. Pribulova, G. Pristas, S. L. Budko, P. C. Canfield, and P. Samuely, Phys. Rev. B \textbf{79}, 012503 (2009).
\bibitem{Mu2009} G. Mu, H. Luo, Z. Wang, L. Shan, C. Ren, and H. H. Wen, Phys. Rev. B \textbf{79}, 174501 (2009).
\bibitem{Terashima2009} K. Terashima, Y. Sekiba, J. H. Bowen, K. Nakayama, T. Kawahara, T. Sato, P. Richard, Y.-M. Xu, L. J. Li, G. H. Cao, Z.-A. Xu,
H. Ding, and T. Takahashi, Proc. Natl. Acad. Sci. USA \textbf{106}, 7330
(2009).
\bibitem{Sanna2009} S. Sanna, R. De Renzi, T. Shiroka, G. Lamura, G. Prando, P. Carretta, M. Putti, A. Martinelli, M. R. Cimberle, M. Tropeano, and A. Palenzona, Phys. Rev. B \textbf{82}, 060508(R) (2010).
\bibitem{Laplace2009} Y. Laplace J. Bobroff, F. Rullier-Albenque, D. Colson, and A. Forget, Phys. Rev. B \textbf{80}, 140501(R) (2009).

\bibitem{Fang2011} C. Fang, Y. Wu, R. Thomale, B. Andrei Bernevig, and J. Hu, Phys. Rev. X \textbf{1}, 011009 (2011).
\bibitem{Mazin2008} I. I. Mazin, D. J. Singh, M. D. Johannes, and M. H. Du, Phys. Rev. Lett. \textbf{101}, 057003 (2008).
\bibitem{Mazin2008_2} I. I. Mazin, M. D. Johannes, L. Boeri, K. Koepernik, D. J. Singh, Phys. Rev. B \textbf{78}, 085104 (2008).
\bibitem{Blatter1994} G. Blatter, M. Y. Feigel'man, Y. B. Geshkenbein, A. I. Larkin, V. M. Vinokur, Rev. Mod. Phys. \textbf{66}, 1125 (1994).
\bibitem{Ni2009} N. Ni, A. Thaler, A. Kracher, J. Q. Yan, S. L. Bud'ko, and P. C. Canfield, Phys. Rev. B \textbf{80}, 024511 (2009).
\bibitem{Ni2008} N. Ni, M. E. Tillman, J.-Q. Yan, A. Kracher, S. T. Hannahs, S. L. Bud'ko, and P. C. Canfield, Phys. Rev. B \textbf{78}, 214515 (2008).

\bibitem{Abrikosov1957} A. A. Abrikosov, Soviet Physics JETP \textbf{5}, 1175 (1957).
\bibitem{Gammel} P. Gammel, Nature \textbf{411}, 434-435 (2001).
\bibitem{Carretta1992} P. Carretta, Phys. Rev. B(R) \textbf{45}, 5760 (1992).
\bibitem{Rigamonti1998} A. Rigamonti, F. Borsa, P.Carretta, Rep. Prog. Phys. \textbf{61}, 1367-1439 (1998).
\bibitem{Carretta1993} P. Carretta, Phys. Rev. B \textbf{48}, 528 (1993).
\bibitem{Torchetti} D. A. Torchetti, M. Fu, D. C. Christensen, K. J. Nelson, T. Imai, H. C. Lei, and C. Petrovic, Phys. Rev. B  \textbf{83}, 104508 (2011).

\bibitem{Prando2011} G. Prando, P. Carretta, R. De Renzi, S. Sanna, A. Palenzona, M. Putti and M. Tropeano, Phys. Rev. B \textbf{83}, 174514 (2011).
\bibitem{Anderson} P. W. Anderson and Y. B. Kim, Rev. Mod. Phys. \textbf{36}, 39 (1964).

\bibitem{Tinkham} M. Tinkham, \textit{Introduction to Superconductivity}, Second Edition,  Dover (1996).
\bibitem{Lascialfari} A. Lascialfari, A. Rigamonti, E. Bernardi, M. Corti, A. Gauzzi and J. C. Villigier, Phys. Rev. B \textbf{80}, 104505 (2009).

\bibitem{Fukazawa2008} H. Fukazawa, K. Hirayama, K. Kondo, T. Yamazaki, Y. Kohori, N. Takeshita, K. Miyazawa, H. Kito, H. Eisaki, A. Iyo, J. Phys. Soc. Jpn. \textbf{77}, 105004 (2008). 
\bibitem{Simmons1962} W. W. Simmons, W. J. O'Sullivan and W. A. Robinson, Phys. Rev. \textbf{127}, 1168 (1962).
\bibitem{Slicther} C. P. Slichter, \textit{Principles of Magnetic Resonance}, Third Enlarged and Updated Edition, Springer Series In Solid-State Sciences 1 (1996).
\bibitem{Oh2011} S. Oh, A. M. Mounce, S. Mukhopadhyay, W. P. Halperin, A. B. Vorontsov, S. L. Bud'ko, P. C. Canfield, Y. Furukawa, A. P. Reyes, and P. L. Kuhns, Phys. Rev. B \textbf{83}, 214501 (2011).
\bibitem{Alloul} H. Alloul, J. Bobroff, and M. Gabay and P. J. Hirschfeld, Rev. Mod. Phys. \textbf{81}, 45 (2009).
\bibitem{Yosida} K. Yosida,  Phys. Rev. \textbf{110}, 769 (1958).
\bibitem{Mac1976} D. E. MacLaughlin, \textit{Solid State Physics: Advances in research and applications}, vol. 3, Academic Press INC., London (1976).
\bibitem{Ning2008} F. Ning, K. Ahailan, T. Imai, A. S. Sefat, R. Jin, M. A. McGuire, B. C. Sales, and D. Mandrus, Jour. Phys. Soc. of Jpn.  \textbf{77}, 103705 (2008).
\bibitem{HS1959} L. C. Hebel and C. P. Slichter, Phys. Rev. \textbf{113}, 1504 (1959).
\bibitem{Parker2008} D. Parker, O. V. Dolgov, M. M. Korshunov, A. A. Golubov,and I. I. Mazin, Phys. Rev. B \textbf{78}, 134524 (2008).
\bibitem{Nakayama2009} K. Nakayama, T. Sato, P. Richard, Y.-M. Xu, Y. Sekiba1, S. Souma, G. F. Chen, J. L. Luo, N. L. Wang, H. Ding and T. Takahashi, Eur. Phys. Lett. \textbf{85} 67002 (2009).

\bibitem{Suh1996} B. J. Suh, F. Borsa, J. Sok, D. R. Torgeson, M. Corti, A. Rigamonti, and Q. Xiong, Phys. Rev. Lett. \textbf{76}, 1928 (1996).
\bibitem{Corti1996} M. Corti, J. Suh, F. Tabak, A. Rigamonti, F. Borsa, M. Xu, B. Dabrowski, Phys. Rev. B \textbf{54}, 9469 (1996).
\bibitem{Brandt1991} E. H. Brandt, Physica B \textbf{169}, 91-98 (1991).
\bibitem{Rotter2008} M. Rotter, M. Tegel, D. Johrendt, I. Schellenberg, W.Hermes, and R. Pottgen, Phys. Rev. B \textbf{78}, 020503(R) (2008).
\bibitem{Pavuna} M. Cyrot and D. Pavuna, \textit{Introduction to Superconductivity and High-Tc materials}, World Scientific Publishing, Singapore (1995).

\bibitem{Brandt1992} H. E. Brandt, Physica C \textbf{195}, 1 (1992).


\bibitem{Clem1991} J. R. Clem, Phys. Rev. B \textbf{43}, 7837 (1991).

\bibitem{Tanatar2009} M. A. Tanatar, N. Ni, C. Martin, R. T. Gordon, H. Kim,V. G. Kogan, G. D. Samolyuk, S. L. Bud'ko, P. C. Canfield, and R. Prozorov, Phys. Rev. B \textbf{79}, 94507 (2009).
\bibitem{Suh1993} B. J. Suh, D. R. Torgeson, F. Borsa, Phys. Rev. Lett. \textbf{71}, 3011 (1993).
\bibitem{Recchia} C. H. Recchia, J. A. Martindale, C. H. Pennington, W. L. Hults and J. L. Smith, Phys. Rev. Lett. \textbf{78}, 3543 (1997).
\bibitem{Sutirtha} S. Mukhopadhyay, S. Oh, A. M.Mounce, M. Lee, W. P. Halperin, N. Ni, S. L. Bud'ko, P.C. Canfield, A. P. Reyes and P. L. Kuhns, New Jour. of Phys. \textbf{11}, 055002 (2009).
\bibitem{Takigawa1986} M. Takigawa and G. Saito, J. Phys. Soc. Jpn. \textbf{55}, 1233 (1986).
\bibitem{Abragam} A. Abragam, \textit{The principles of Nuclear Magnetism}, Oxford Univesity Press (1961).
\bibitem{Williams2010} T. J. Williams, A. A. Aczel, E. Baggio-Saitovitch, S. L. Bud'ko, P. C. Canfield, J. P. Carlo, T. Goko, H. Kageyama, A. Kitada, J. Munevar, N. Ni, S. R. Saha, K. Kirschenbaum, J. Paglione, D. R. Sanchez-Candela, Y. J. Uemura and G. M. Luke, Phys. Rev. B \textbf{82}, 094512 (2010).
\bibitem{Gordon2009} R. T. Gordon, N. Ni, C . Martin, M. A. Tanatar, M.D. Vannette, H. Kim, G. D. Samolyuk, J. Schmalian, S. Nandi, A. Kreyssig, A. I. Goldman, J. Q. Yan, S. L. Budko, P. C. Canfield, and R. Prozorov, Phys. Rev. Lett.\textbf{102}, 127004 (2009).
\bibitem{Brandt1988} E. H. Brandt, Phys. Rev. B(R) \textbf{37}, 2349 (1988).






\end{references}
\end{document}